\documentclass[doublecol]{epl2} 
\usepackage{latexsym}
\usepackage{graphicx}
\usepackage{rotating}
\usepackage{longtable}
\usepackage{multirow}

\title{Stress engineering at the nanometer scale: two-component adlayer stripes}

\author{T.~O.~Mente\c{s}\inst{1} \and N.~Stoji\'{c}\inst{2,3} \and A.~Locatelli\inst{1} \and 
L.~Aballe\inst{4} \and N.~Binggeli\inst{2,5} \and M.~A.~Ni\~{n}o\inst{1} \and 
M.~Kiskinova\inst{1} \and E.~Bauer\inst{6}}

\institute{
 \inst{1} Sincrotrone Trieste S.C.p.A., Basovizza-Trieste 34149, Italy\\
 \inst{2} Abdus Salam International Centre for Theoretical Physics, Strada Costiera 11, Trieste 34151, Italy\\
 \inst{3} IOM-CNR Democritos, Theory @ Elettra group,  Trieste 34151, Italy\\
 \inst{4} CELLS-ALBA, Carretera BP 1413, km 3.3, 08290 Cerdanyola del Vall\'{e}s, Barcelona, Spain\\
 \inst{5} IOM-CNR Democritos,  Trieste 34151, Italy\\
 \inst{6} Department of Physics, Arizona State University, Tempe, Arizona 85287-1504, USA
}


\abstract{Spontaneously-formed equilibrium nanopatterns with long-range order are
widely observed in a variety of systems, but their pronounced temperature dependence
remains an impediment to maintain such patterns away from the temperature of formation.
Here, we report on a highly-ordered stress-induced stripe pattern 
in a two-component, Pd-O, adsorbate monolayer on W(110),
produced at high temperature and identically preserved at lower temperatures.
The pattern shows a tunable period (down to 16~nm) and orientation, 
as predicted by a continuum model theory along with the
surface stress and its anisotropy found in our DFT calculations. 
The control over thermal fluctuations in the stripe formation process is based on 
the breaking/restoring of ergodicity in a high-density lattice gas with long-range interactions 
upon turning off/on particle exchange with a heat bath.}

\begin{document}

\maketitle


Spontaneous formation of periodic patterns is an example of nature's tendency towards order.
A class of such structures is induced by surface stress~\cite{ibach97}, 
and has been widely observed on single crystal 
surfaces~\cite{kerniesch91, jonpelhon96, hanbarswa97, pohbarfig99, plalasbar01, vanplabar03, figleobar08}.
It is well-known that the formation of these equilibrium patterns 
is driven by a competition between interactions 
at different length scales~\cite{alevanmea88}. 
The forces in action, due to short-range near-neighbour and long-range dipolar interactions, 
are of the most general type resulting in very similar phenomena occuring also in
magnetic~\cite{debmacwhi00} and electrostatic systems~\cite{andbrojoa87}. 
This gives a strong motivation to include such equilibrium phases with mesoscopic order
in the discussion of making self-assembly available to future applications~\cite{whigrz2002}.

To this date, the majority of the reports on stress-induced equilibrium patterns have focused on
the characteristic feature shape and size. The temperature dependence of the pattern period
appears as a common ingredient in the various systems studied, and the general explanation is
based on a scaling of the interaction parameters due to thermal disorder~\cite{menlocaba08}. 
The pronounced thermal fluctuations~\cite{pohbarfig99, vanplabar03}, derived from the high mobility
of the adatoms, are essential in reaching the thermodynamic equilibrium phase.
However,  this mobility, which allows the pattern to almost instantaneously equilibrate, 
is a hindrance to preserving the pattern away from its formation temperature.

\begin{figure*}[ht]
\begin{center}
\onefigure{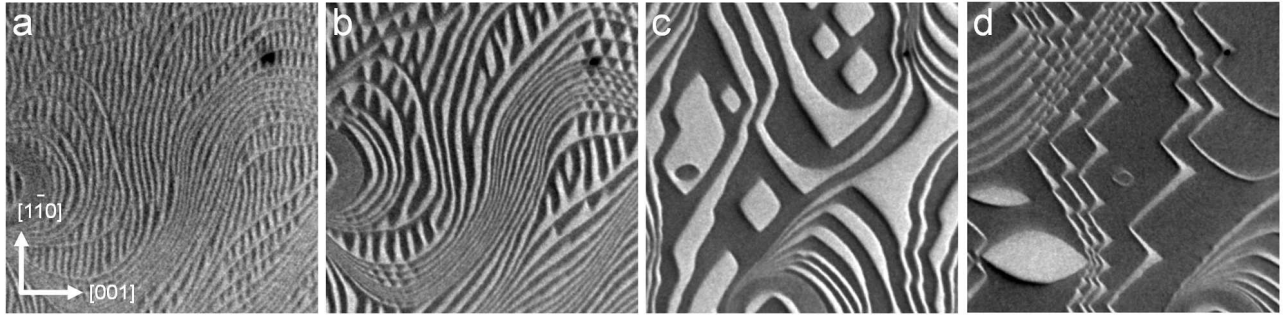}
\caption{Oxygen adsorption on 0.4 ML of Pd on W(110) at constant temperature (about 1200~K). 
Panels (a) through (d) are snapshots at oxygen doses 0.54, 0.68, 1.26 and 
3.51 Langmuir, respectively. The corresponding oxygen coverages are 
0.08, 0.10, 0.19 and 0.40 ML. At 5~eV electron energy, bright areas correspond to the
Pd monolayer. Curved lines are W single atomic steps. The image size is 3.5~$\mu m$ $\times$ 3.5~$\mu m$.}
\label{fig:dosing}
\end{center}
\end{figure*}

Here, we explore the possibility of controlling the thermal fluctuations
by adding a second adspecies to an already stripe-forming system.  The idea is based on the
slow dynamics in a high-density binary lattice gas, which leads to a tendency towards glassy behaviour.
In particular, we implement this by dosing small amounts of oxygen 
on submonolayer Pd/W(110). The oxygen-free surface is known to organize 
at high temperatures into alternating stripes of a dense Pd monolayer and 
a dilute Pd lattice-gas~\cite{menlocaba08}.
In the following, it will be shown that the mixed Pd-O layer necessitates 
special growth conditions in order to form a similar stripe phase. 
As we will describe shortly, this allows to identically preserve the equilibrium pattern, 
produced at high temperature, at lower temperatures.
Moreover, the addition of oxygen leads to a progressive change in
the surface stress, which is reflected by the changes in pattern period and anisotropy.
The result is a highly-ordered stripe pattern
with a tunable period down to below 20~nm, 
and with an orientation that can be rotated by 90$^\circ$.
The density-functional theory (DFT) calculations, in conjunction with 
continuum model results, are used to predict the phase diagram of stripe 
shape and orientation of this system as a function of Pd and oxygen 
coverages. This allows us to identify the equilibrium phases and 
successfully account for the main observed changes in stripe pattern, 
including the 90$^\circ$ rotation in the stripe orientation.

Using low-energy electron microscopy (LEEM)~\cite{bauer94,altman10,locabamen06}
we studied Pd-O pattern formation
on W(110) following different growth pathways and temperatures.
The progressive changes to the Pd/W(110) stripes upon oxygen dosing and 
the kinetic factors in forming the new Pd-O stripe pattern are revealed in the two approaches
to be summarized,
i) dosing oxygen on a partially Pd-covered surface at 1200~K, and 
ii) depositing Pd on a partially oxygen-covered surface at 1240~K. 
The latter results in the Pd-O stripe phase 
as Pd is deposited at a temperature at which it partially desorbs. 
This new stripe phase, with perpendicular stripe orientation relative to 
the Pd stripes on the oxygen-free surface, is identified as an equilibrium configuration.
The conditions under which the Pd-O stripes are obtained involve effective 
fluctuations via partial desorption of Pd, which
enable to reach the minimum energy configuration.

Oxygen on W(110) forms a series of ordered phases as a
function of coverage~\cite{johwilchi93} and has been
extensively studied as a model system for
surface diffusion and island formation~\cite{wutrilag89}.
For coverages below 0.5~ML, relevant to this study, it 
is found in (1$\times$2)-ordered islands.
Importantly, oxygen has a strong tendency to laterally segregate from most transition
metal adatoms coadsorbed on W(110)~\cite{nahgom97}.
Furthermore, recent work has shown that oxygen induces a considerable stress
change on tungsten~\cite{menstobin08}. These observations 
point towards the possibility of finding an ordered two-component (oxygen and metal) adsorbate phase.

Following this reasoning, we first dosed molecular oxygen on a partially
Pd-covered W(110) surface at 1200~K, above the stripe disordering temperature.
At this temperature, Pd shows only short-range order~\cite{menlocaba08}.
Upon adsorbing small amounts of oxygen the stripe phase appears along $[1\bar{1}0]$ 
with a wavy morphology as shown in fig.~\ref{fig:dosing}a. 
With increasing O coverage the pattern period increases (fig.~\ref{fig:dosing}b),
followed by segregation
of Pd into large 2D islands surrounded by O (fig.~\ref{fig:dosing}c).
Further increase in O coverage (above roughly 0.3 ML) results in the
shrinking of the Pd islands (fig.~\ref{fig:dosing}d). 
This suggests that when the local density of O
exceeds that of the (1$\times$2)-ordered layer, 
Pd atoms are displaced into the second layer, 
from which they readily desorb at this high
temperature. It is important to note that at this temperature the order of deposition
has little influence on the final result. In other words, depositing Pd on a partially 
oxygen precovered surface below 1200~K
leads to similar structures as displayed in fig.~\ref{fig:dosing}d.

\begin{figure}[ht]
\begin{center}
\onefigure{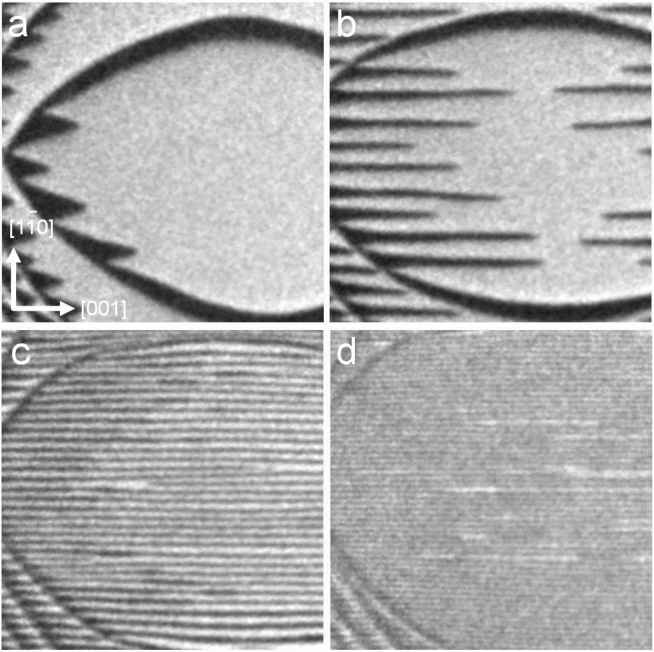}
\caption{Pd adsorption on 0.33 ML of O on W(110) at 1240~K. 
Panels (a) through (d) are snapshots at Pd coverages 0.18, 0.23, 0.27 and 
0.31~ML, respectively. The stripe direction corresponds to the substrate $[001]$ axis.
At 13.5~eV electron energy, dark areas correspond 
to regions covered with Pd. 
W single atomic steps appear as curved lines.
The image size is 1.0~$\mu m$ $\times$ 1.0~$\mu m$.}
\label{fig:adsorption}
\end{center}
\end{figure}

In the second approach, 
Pd was adsorbed at 1240~K
(about 40 degrees higher compared to that in fig.~\ref{fig:dosing}) on a surface 
precovered with less than 0.5~ML of oxygen. This temperature
corresponds to the tail of the thermal desorption peak of 
the first Pd adlayer~\cite{schbau1980}, 
so that Pd adsorption is partially balanced by desorption. 
Under these conditions, a highly-ordered stripe pattern
develops along the [$001$] direction as
illustrated in fig.~\ref{fig:adsorption}.
Crucially, partial desorption was established as the necessary condition for forming the
mesoscopic pattern.
The stripe pattern consists of alternating Pd and O phases.
The stripe period strongly depends on the amount of Pd on the surface, 
and a minimum period of about $16$~nm is reached
at the highest Pd coverage (about 0.33~ML)
allowed by the preadsorbed O.
The new Pd-O stripe phase differs from the pure Pd stripes
in the orientation of the stripes ($[001]$ vs. $[1\bar{1}0]$), 
i.e. by a 90$^\circ$ rotation in the stripe direction,
as well as in the improved directionality and reduced periodicity (16 vs. 60~nm).
Moreover, the pattern can be frozen upon cooling and thus preserved at room temperature.

In order to elucidate whether this new phase represents the thermodynamic equilibrium,
the energetics and anisotropy of the Pd-O adsorption system 
have been investigated by DFT calculations,
providing the parameters to the continuum-model arguments developed for stress-induced
equilibrium patterns~\cite{gaolusuo2002,lusuo2002,gaosuo2003}.
The DFT pseudopotential calculations were performed in a plane-wave basis by using the
PWscf code~\cite{giabarbon09}. 
The details of the surface stress calculations for W and O/W can be found
in Refs.~\cite{menstobin08} and~\cite{stomenbin10}.
The Pd-O boundary energy was calculated on a 6-layer slab
consisting of a Pd-O surface layer and 5~W layers. The lateral size of the slab corresponds to
10$\times$2 substrate unit cells, of which one half
is covered by Pd and the other half by (1$\times$2)O. 
In estimating the creation energy of a single boundary,
only the first atoms adjacent to the Pd-O($1\times2$) boundary were allowed to relax (in the upper
4 layers) in order to minimize contribution from
elastic interactions between neighbouring boundaries.

It should be noted that the atomic structures of
both Pd and O stripes at high temperature are
a challenge for modeling.
On one hand, the low-temperature ordered oxygen phases
lose their long-range order several hundred degrees below the temperature at which stripes form
(for example the ($1\times2$) structure disorders at about 700~K~\cite{wutrilag89}).
However, the short-range order persists at high temperatures, 
and influences the stripe phase profoundly. In particular, 
a local ($1\times2$) density is the highest oxygen packing that can be attained in the presence of
Pd islands, before Pd is displaced to the second and higher layers.
Moreover, the oxygen stripe in the Pd-O stripe phase has the ($1\times2$) structure when 
cooled down to room temperature. Therefore, 
at high temperature the oxygen-covered regions
can be described as having short-range ($1\times2$) order. 
The effect of disorder on the oxygen surface stress is studied in Ref.~\cite{stomenbin10}, and
the values obtained are reproduced in Table~\ref{tab:dft_stress}.

\renewcommand{\baselinestretch}{1}
\begin{table}[ht]
\bigskip
\begin{center}
\begin{tabular}{ c c c c}
\hline
\hline \\[-2mm]
       \multirow{2}{*}{direction}
       & $\tau_{W(1\times1)}$
       & $\tau^{high T}_{O(1\times2)}$
       & $\tau_{Pd(5\%mismatch)}$ \\ [2mm]
\hline  \\[-3mm]
 $[1\bar{1}0] $   & $3.6$    &  $-0.2$     & $3.1$     \\
 $[001]         $   & $5.2$     &  $3.4$     &  $3.4$     \\[2mm]
\hline
\hline  \\[-3mm]
       & $\Delta\tau_{Pd-W}$
       & $\Delta\tau_{Pd-O}$
       & $\beta_{Pd-O}$ \\ [2mm]
\hline  \\[-3mm]
 $[1\bar{1}0] $   & $-0.5$    &  $3.3$     &  $0.152$   \\
 $[001]         $   & $-1.8$     &  $\sim0$     &  $0.164$    \\[2mm]
\hline
\hline
\end{tabular}
\caption{Calculated surface stresses ($\tau$) and the boundary energy ($\beta$) 
between Pd and O phases on W(110). The upper two rows list the stress values. 
The lower two rows display the directional dependence of stress differences and 
boundary energy. All stresses are given in $N/m$, and the boundary energy 
has units eV/\AA. O$(1\times2)$ stress is taken from Ref.~\cite{stomenbin10}.}
\label{tab:dft_stress}
\end{center}
\end{table}

On the other hand, at high temperature the
submonolayer Pd is pseudomorphic on W(110) only along  $[001]$ and 
shows a mismatch structure along $[1\bar{1}0]$ (i.e. one-dimensional pseudomorphism)~\cite{schbau1980}. 
Using low-energy electron diffraction, we observe a very similar situation for Pd within the Pd-O stripes 
with a mismatch of approximately 5~\%.
We have estimated the effect of this non-pseudomorphism on the surface stress by
varying the mean lattice constant of the Pd overlayer along  $[1\bar{1}0]$
for selected commensurate periodicities 
in DFT calculations using supercells with large $[1\bar{1}0]$ dimensions
on asymmetric slabs~\cite{note_stojic}. 
The calculated values for 5~\% mismatch are displayed in the upper half of Table~\ref{tab:dft_stress}.
Note that small variations of the mismatch (less than 1~\%) due to changes in temperature or coverage
are found to modify the calculated stresses only slightly and
have no consequence in the following discussion.

\begin{figure*}[t]
\begin{center}
\onefigure[scale=0.95]{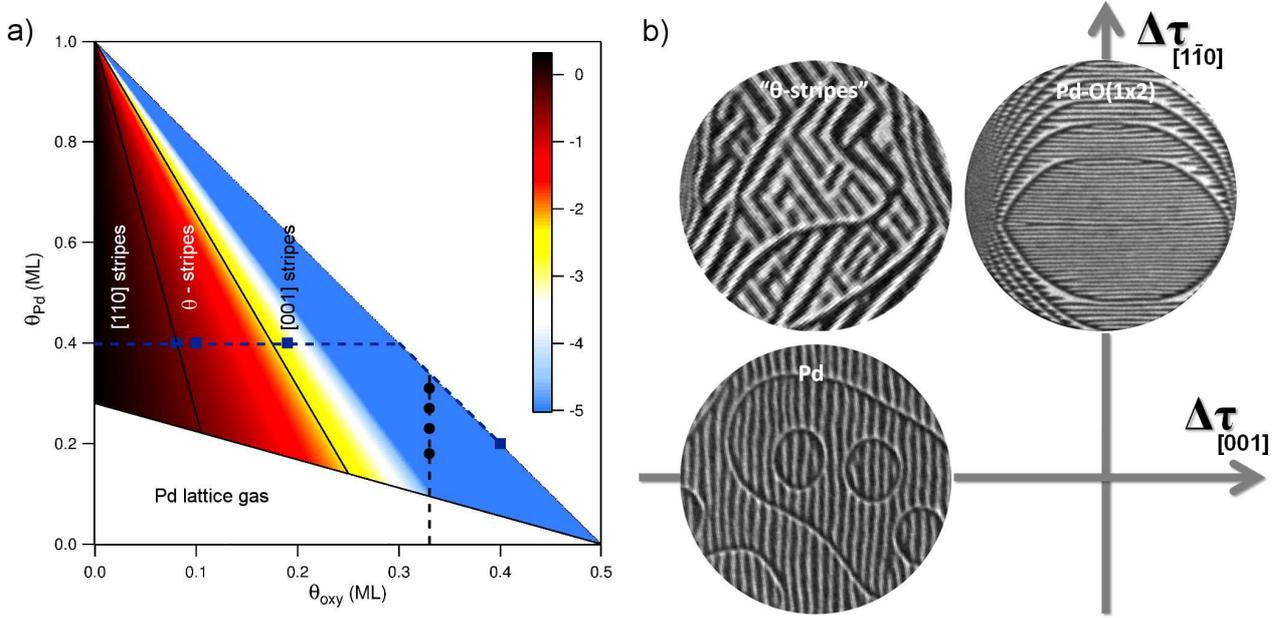}
\caption{(Color online) a) The phase diagram as a function of Pd and oxygen 
coverages generated from the calculated stresses.
The color scale gives the $\Delta \tau_{[1\bar{1}0]} / \Delta \tau_{[001]}$ ratio.
Solid diagonal lines are contours corresponding to the critical ratio, $-0.44$ and $-1/0.44$. 
Blue and black dashed lines trace the coverages in fig.~\ref{fig:dosing} and 
fig.~\ref{fig:adsorption}, respectively. Solid squares and circles mark the panels in each figure.
Total coverage is limited to below a pseudomorphic monolayer. 
The Pd lattice gas is sketched for temperatures near Pd disordering.
b) Three distinct adsorbate patterns on W(110). The patterns are qualitatively sketched on a diagram
of stress change across the stripe boundary along the two crystallographic directions. All images 
have 2.0~$\mu m$ diameter.}
\label{fig:labyrinth}
\end{center}
\end{figure*}

The results of the calculations relevant to the stripe formation, i.e. the directional dependence of
stress difference and boundary energy between the two phases (O and Pd),
are summarized in the last two rows in Table~\ref{tab:dft_stress}. For comparison, we
include the stress anisotropy for the oxygen-free Pd stripes in the first column.
An important finding is that the Pd-O boundary energy is nearly isotropic.
The values in the table clearly identify the stress difference as
the real source of anisotropy.
The elastic energy gain is higher for a larger
stress difference across the stripe boundary, 
i.e. perpendicular to the stripe direction. As a consequence, the expected stripe direction strikingly changes
from $[1\bar{1}0]$ for oxygen-free Pd stripes to $[001]$ for Pd-O stripes on W(110).

The calculated stress anisotropy is found to be more than sufficient for 
driving the formation of the Pd-O stripes along $[001]$ according to the phase diagram 
(see fig.~\ref{fig:labyrinth}a) evaluated on the basis of a continuum-model 
described by Suo and coworkers~\cite{gaolusuo2002,lusuo2002,gaosuo2003} and our
computed stresses. The model predicts distinct stripe phases as the ratio of stress 
differences along the two high-symmetry crystallographic directions,
$r=\Delta \tau_{[1\bar{1}0]} / \Delta \tau_{[001]}$, is varied.
A critical value, $r_c = 1+2\nu$ (where $\nu$ is the Poisson's ratio), is found analytically by minimizing the free
energy as a function of stripe orientation, which marks the onset of regular stripes~\cite{gaolusuo2002}.
In particular for an elastically isotropic substrate as W(110)~\cite{leobarkel2005},
when $r$ is greater than $r_c = -0.44$ 
(or smaller than $-1/0.44$ for the perpendicular stripes) 
a linear stripe phase is expected. 
As can be seen from the results in Table~\ref{tab:dft_stress}, 
this condition is satisfied both for Pd and Pd-O stripes.
In order to check this quantitative criterion for all coverages displayed 
in fig.~\ref{fig:adsorption}, the stress differences can be estimated by subtracting a weighted
sum of oxygen and tungsten surface stresses from that of Pd. 
This is based on the observation that increasing Pd coverage squeezes 
the oxygen atoms towards a fully-packed ($1\times2$) order at the highest density.
The corresponding stress ratios, shown as dark circles in fig.~\ref{fig:labyrinth}a,
confirm the presence of $[001]$ stripes for 0.33~ML O coverage, regardless of the Pd coverage.

Based on the continuum-model, a herringbone pattern 
(``$\theta$-stripes"~\cite{gaolusuo2002}) is expected when the stress differences 
are of similar magnitude but of opposite sign along orthogonal directions,
with the $\Delta\tau$-ratio in the range $-1/0.44<r<-0.44$.
Inspecting Table~\ref{tab:dft_stress}, this corresponds to an intermediate situation
between O-free Pd stripes and fully packed Pd-O stripes. In agreement with this prediction,
the herringbone pattern, 
already hinted by the wavy stripes shown in fig.~\ref{fig:dosing}b, is observed
in perfect form when the experiment is repeated, always under simultaneous Pd adsorption-desorption, 
with much lower oxygen coverage. 
The resulting pattern is displayed in fig.~\ref{fig:labyrinth}b, 
together with the O-free Pd and Pd-O stripe phases.  
These three patterns correspond to the three different solid-state phases of the overlayer 
in the phase diagram of fig.~\ref{fig:labyrinth}a.

\begin{figure}[t]
\begin{center}
\onefigure[scale=0.9]{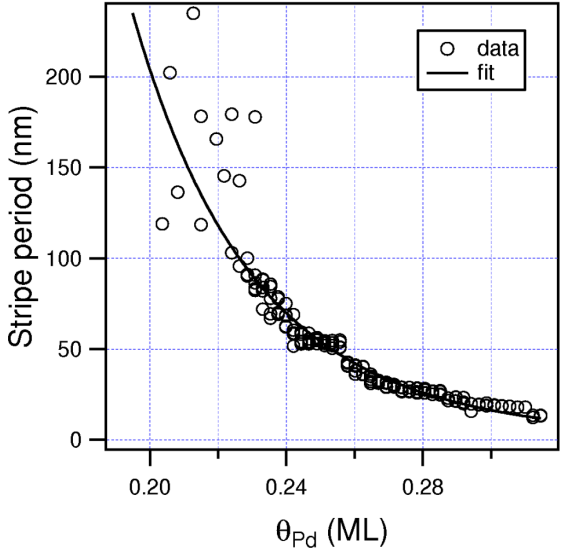}
\caption{Stripe period as a function of Pd coverage.  The fit function is
described in the text. }
\label{fig:period}
\end{center}
\end{figure}

In addition to the pattern shape and orientation, the stripe period shows a strong
coverage dependence. Using the expression from the continuum model~\cite{alevanmea88},
the period, $D$, is given by,
\begin{equation}
D(\theta_{Pd},\theta_{O})\ =\ 
      \frac{A\ e^{\frac{B}{\Delta\tau(\theta_{Pd},\theta_{O})^2}}}{cos(\frac{\pi(1-2\theta_{Pd})}{2})} 
\end{equation}
where $\theta_{Pd}$ and $\theta_{O}$ are the Pd and O coverages,
$\Delta\tau$ is the coverage-dependent stress difference 
between the fully-packed Pd and the partially-covered oxygen stripes,
A and B are the fitting parameters corresponding to the effective boundary width and a combination
of energetic factors, respectively.
The denominator accounts for the effect of stripes with unequal widths~\cite{alevanmea88}.
The stress difference, $\Delta\tau$, is found by a weighted average of the
calculated values displayed in Table~\ref{tab:dft_stress} as the Pd coverage is varied. 
The excellent fit to the experimental data is shown in fig.~\ref{fig:period}.
The only fit parameters, A and B, are found to be $0.38 \pm 0.04$~\AA\ and $55.6 \pm 1.0$~$N^2/m^2$,
respectively.

The fact that the experimental stripe period, shape and orientation show very good agreement
with the continuum model (along with the DFT-calculated energy and stress parameters) confirms
that the observed stripes represent the minimum of the free energy, 
which is determined by the competition between short range 
(atomic bonding) and long range (elastic) interactions. 
In other words, the stripes are in thermodynamic equilibrium.
However, as described in the context of the measurements, 
the necessary condition for obtaining this equilibrium phase is
simultaneous Pd adsorption-desorption, i.e. an open system. 
On the other hand, in the closed Pd-O system
the totally segregated state seen in fig.~\ref{fig:dosing}c is found, 
which is apparently a local minimum of the free energy.
The kinetic barriers in this complex free-energy landscape are overcome only
with increased fluctuations via exchange with a heat-bath (3D gas), 
allowing the system to reach the real equilibrium configuration, i.e. the stripe pattern.

It is to be expected that a
system with long-range interactions should face heightened kinetic 
barriers in its search for the global minimum in the configuration space. The basis for this statement 
is apparent in the formation of the equilibrium state. 
Generally, the predominant mechanism in the
decay towards the minimum energy phase is by nucleation of spatially confined {\em droplets}, 
which progressively increase in size~\cite{binder87}.
Such a possibility does not exist
for a phase stabilized by long-range forces, which necessitate
its formation in a coherent manner within spatially extended regions. 
The difference is clear in studies of dense and viscous lattice gas models. In particular,
systems with short-range interactions alone do not show glassy behaviour, in spite of their slow-dynamics
at high density~\cite{fusgalpet02}. On the other hand, long-range interactions in a thermodynamically closed
system may lead to broken ergodicity and infinite-lifetime metastable states~\cite{mukrufsch05}. 

Importantly, it has recently been shown that the ergodicity, which is broken in a closed system due to
long-range interactions, is restored by coupling it to a heat-bath~\cite{balchaorl09}. 
This distinction between thermodynamically closed (microcanonical) and open (canonical) 
systems in their decay to the thermal equilibrium state provides 
a direct explanation to our observations~\cite{note_thermodynamics}.
In other words, the simultaneous Pd adsorption-desorption condition is identified as coupling 
the otherwise-closed two-dimensional adsorbate layer to a Pd reservoir (three-dimensional Pd gas), 
which acts as a heat-bath and allows the formation of the equilibrium stripe phase.

In conclusion, we have shown  the presence of a stress-induced
Pd-O stripe phase on W(110). The anisotropy and the period of the stripes depend sensitively 
on the surface stress, and can be tuned by changing the relative coverage of the two species.
Due to the presence of long-range interactions, the high-density lattice gas is
able to decay into the equilibrium stripe phase only via material exchange with the gas phase.
As a consequence of suppressed diffusivity, the pattern can be preserved by 
cooling to lower temperatures. 
This opens a new path towards fabricating highly-ordered periodic structures 
via self-organization at the nanometer scale.



\begin{thebibliography}{10}
\expandafter\ifx\csname url\endcsname\relax\def\url#1{\texttt{#1}}\fi

\bibitem{ibach97}
\Name{Ibach H.}
  \REVIEW{Surf.\ Sci.\ Rep. }{29}{1997}{195}.

\bibitem{kerniesch91}
\Name{Kern K., Niehus H., Schatz A., Zeppenfeld P., Goerge J. \and Comsa G.}
  \REVIEW{Phys.\ Rev.\ Lett. }{67}{1991}{855}.

\bibitem{jonpelhon96}
\Name{Jones D.~E., Pelz J.~P., Hong Y., Bauer E. \and Tsong I. S.~T.}
  \REVIEW{Phys.\ Rev.\ Lett. }{77}{1996}{330}.

\bibitem{hanbarswa97}
\Name{Hannon J.~B., Bartelt N.~C., Swartzentruber B.~S., Hamilton J.~C. \and
  Kellogg G.~L.} \REVIEW{Phys.\ Rev.\ Lett. }{79}{1997}{4226}.

\bibitem{pohbarfig99}
\Name{Pohl K., Bartelt M.~C., de~la Figuera J., Bartelt N.~C., Hrbek J. \and
  Hwang R.~Q.} \REVIEW{Nature }{397}{1999}{238}.

\bibitem{plalasbar01}
\Name{Plass R., Last J., Bartelt N.~C. \and Kellogg G.~L.} \REVIEW{Nature
  }{412}{2001}{875}.

\bibitem{vanplabar03}
\Name{van Gastel R., Plass R., Bartelt N.~C. \and Kellogg G.~L.} \REVIEW{Phys.\
  Rev.\ Lett. }{91}{2003}{055503}.

\bibitem{figleobar08}
\Name{de~la Figuera J., L\'{e}onard F., Bartelt N.~C., Stumpf R. \and McCarty
  K.~F.} \REVIEW{Phys.\ Rev.\ Lett. }{100}{2008}{186102}.

\bibitem{alevanmea88}
\Name{Alerhand O.~L., Vanderbilt D., Meade R.~D. \and Joannopoulos J.~D.}
  \REVIEW{Phys.\ Rev.\ Lett. }{61}{1988}{1973}.

\bibitem{debmacwhi00}
\Name{De'Bell K., MacIsaac A.~B. \and Whitehead J.~P.} \REVIEW{Rev.\ Mod.\
  Phys. }{72}{2000}{225}.

\bibitem{andbrojoa87}
\Name{Andelman D., Bro\c{c}hard F. \and Joanny J.-F.} \REVIEW{J.\ Chem.\ Phys.
  }{86}{1987}{3673}.

\bibitem{whigrz2002}
\Name{Whitesides G.~M. \and Grzybowski B.} \REVIEW{Science }{295}{2002}{2418}.

\bibitem{menlocaba08}
\Name{Mente\c{s} T.~O., Locatelli A., Aballe L. \and Bauer E.} \REVIEW{Phys.\
  Rev.\ Lett. }{101}{2008}{085701}.

\bibitem{bauer94}
\Name{Bauer E.} \REVIEW{Rep.\ Prog.\ Phys. }{57}{1994}{895}.

\bibitem{altman10}
\Name{Altman M.} \REVIEW{J.\ Phys.:\ Condens. Matter }{22}{2010}{084017}.

\bibitem{locabamen06}
\Name{Locatelli A., Aballe L., Mentes T.~O., Kiskinova M. \and Bauer E.}
  \REVIEW{Surf.\ Interface\ Anal. }{38}{2006}{1554}.

\bibitem{johwilchi93}
\Name{Johnson K.~E., Wilson R.~J. \and S.Chiang} \REVIEW{Phys.\ Rev.\ Lett.
  }{71}{1993}{1055}.

\bibitem{wutrilag89}
\Name{Wu P.~K., Tringides M.~C. \and Lagally M.~G.} \REVIEW{Phys.\ Rev.\ B
  }{39}{1989}{7595}.

\bibitem{nahgom97}
\Name{Nahm T.-U. \and Gomer R.} \REVIEW{Surf.\ Sci. }{373}{1997}{237}.

\bibitem{menstobin08}
\Name{Mente\c{s} T.~O., Stoji\'{c} N., Binggeli N., Ni{\~n}o M.~A., Locatelli
  A., Aballe L., Kiskinova M. \and Bauer E.} \REVIEW{Phys.\ Rev.\ B
  }{77}{2008}{155414}.

\bibitem{schbau1980}
\Name{Schlenk W. \and Bauer E.} \REVIEW{Surf.\ Sci. }{93}{1980}{9}.

\bibitem{gaolusuo2002}
\Name{Gao Y.~F., Lu W. \and Suo Z.} \REVIEW{Acta\ Mater. }{50}{2002}{2297}.

\bibitem{lusuo2002}
\Name{Lu W. \and Suo Z.} \REVIEW{Phys.\ Rev.\ B }{65}{2002}{085401}.

\bibitem{gaosuo2003}
\Name{Gao Y.~F. \and Suo Z.} \REVIEW{J.\ Mech.\ Phys.\ Solids }{51}{2003}{147}.

\bibitem{giabarbon09}
\Name{Giannozzi P., Baroni S., Bonini N., Calandra M., Car R., Cavanozzi C.,
  Ceresoli D., Chiarotti G.~L., Cococcioni M., Dabo I. \and et~al.} \REVIEW{J.\
  Phys.:\ Condens.\ Matter }{21}{2009}{395502}.

\bibitem{stomenbin10}
\Name{Stoji\'{c} N., Mente\c{s} T.~O., Binggeli N., Ni{\~n}o M.~A., Locatelli
  A. \and Bauer E.} \REVIEW{Phys.\ Rev.\ B }{81}{2010}{115437}.

\bibitem{note_stojic}
N. Stoji\'{c} {\em et al., in preparation.}

\bibitem{leobarkel2005}
\Name{L\'{e}onard F., Bartelt N.~C. \and Kellogg G.~L.} \REVIEW{Phys.\ Rev.\ B
  }{71}{2005}{045416}.

\bibitem{binder87}
\Name{Binder K.} \REVIEW{Rep.\ Prog.\ Phys. }{50}{1987}{783}.

\bibitem{fusgalpet02}
\Name{Fusco C., Gallo P., Petri A. \and Rovere M.} \REVIEW{Phys.\ Rev.\ E
  }{65}{2002}{026127}.

\bibitem{mukrufsch05}
\Name{Mukamel D., Ruffo S. \and Schreiber N.} \REVIEW{Phys.\ Rev.\ Lett.
  }{95}{2005}{240604}.

\bibitem{balchaorl09}
\Name{Baldavin F., Chavanis P.-H. \and Orlandini E.} \REVIEW{Phys.\ Rev.\ E
  }{79}{2009}{011102}.

\bibitem{note_thermodynamics}
Such a comparison is possible as the extensive nature of energy is preserved
in the presence of the dipolar-type long-range elastic interactions: JUND P., KIM S. G. and TSALLIS C.,
{\em Phys. Rev. B} , {\bf 52} (1995) 50.

\end{thebibliography}

\end{document}